\documentclass[iop,revtex4]{emulateapj}
\slugcomment{{\sc Accepted to MNRAS:} September 6, 2013} 
\shortauthors{Walkowicz et al.}  
\shorttitle{Rotation Periods and Variability for KOIs}
\usepackage[caption=false]{subfig}
\begin{document}

\title{Rotation Periods, Variability Properties and Ages for {\em Kepler} Exoplanet Candidate Host Stars}

\author{
Lucianne M. Walkowicz\altaffilmark{1} and Gibor S. Basri\altaffilmark{2}
}

\altaffiltext{1}{Department of Astrophysical Sciences, Princeton University, 4 Ivy Lane, Princeton NJ 08544}
\altaffiltext{2}{Astronomy Department, University of California at Berkeley, 601 Campbell Hall, Berkeley, CA 94720}

\begin{abstract}

We report rotation periods, variability characteristics, gyrochronological ages for $\sim$950 of the {\em Kepler} Object of Interest host stars.  We find a wide dispersion in the amplitude of the photometric variability as a function of rotation, likely indicating differences in the spot distribution among stars. We use these rotation periods in combination with published spectroscopic measurements of vsini and stellar parameters to derive the stellar inclination in the line-of-sight, and find a number of systems with possible spin-orbit misalignment. We additionally find several systems with close-in planet candidates whose stellar rotation periods are equal to or twice the planetary orbital period, indicative of possible tidal interactions between these planets and their parent stars. If these systems survive validation to become confirmed planets, they will provide important clues to the evolutionary history of these systems. 


\end{abstract}


\section{INTRODUCTION}

Since its launch in the spring of 2009, NASA's {\em Kepler} mission \citep{2010ApJ...713L..79K} has found almost 3000 planet candidates, known as Kepler Objects of Interest \citep[or KOIs;][]{2013ApJS..204...24B}. Although the primary goal of the {\em Kepler} mission is to determine the frequency of Earth-size and larger planets in the habitable zones of Sun-like stars, {\em Kepler}'s high precision photometry has proven to be an invaluable window into stellar astrophysics.  {\em Kepler}'s nearly continuous, rapid cadence monitoring of stars reveals the manifestations of stellar variability to greater precision than previously possible for most stars besides our Sun. 

In the solar-like stars that comprise the majority of the {\em Kepler} exoplanet search targets, much of this variability is due to magnetic activity: starspots and active regions on the stellar surface rotate into and out of view, modulating the stellar brightness on the rotation timescale of the star. The very generation of the magnetic field is intimately tied to the stellar rotation and differential rotation \citep{1975ApJ...198..205P}, and so in observing the variability caused by these surface manifestations of the field, one can measure rotation periods and differential rotation, thereby providing feedback to theory of magnetic field generation in stars both similar to and different than our Sun. While many studies of stellar magnetism have traditionally been limited to high amplitude variability that could easily be measured from ground-based photometry, and to shorter rotation periods that could be derived from less-densely sampled data, {\em Kepler}'s large sample of precision photometry reveals levels of variability that are comparable to that of the star from which most of our knowledge of stellar magnetism is drawn: our own Sun \citep{2010ApJ...713L.155B,2013ApJ...769...37B}.

In the case of exoplanet host stars, magnetic activity can reveal a variety of information about the system. The most often-discussed aspect of stellar influence on attendant planets is whether said planets reside in the ``habitable zone'', the range of distance from the star where the stellar insolation is similar to that of our Sun on Earth, such that liquid water could conceivably exist on the surface of a suitably terrestrial planet \citep{1993Icar..101..108K,2013ApJ...765..131K}. The habitable zone is usually based on the bulk luminosity of the star, rather than anything in particular about its exact spectral energy distribution. As such it often ignores the effects of stellar activity, which causes emission at UV and X-ray wavelengths. This high-energy radiation influences planetary atmospheric composition and evolutionary history by driving photochemistry and atmospheric escape \citep[e.g.][]{1996LPI....27..655K,2010AsBio..10..751S,2008SSRv..139..437Y}. The high energy radiation environment is often more difficult to measure, but as its underlying cause is the same magnetic field that drives the optical variability, it can be inferred to some extent from rotational variability as well as the presence of stellar flares. 

Stellar variability can also provide a means of learning the age, architecture and history of planetary systems. Solar-like stars (those with outer convective layers) arrive on the Main Sequence rotating rapidly, but then spin down with age as they lose angular momentum via braking by magnetized stellar winds, leading to the stellar rotation-activity-age relations \citep[e.g.][]{1972ApJ...171..565S,1988ApJ...334..436B,1995ApJ...438..269B}. The monotonic relationship between stellar rotation and age on the Main Sequence has lead to the development of so-called ``gyrochronology'' relations \citep{2007ApJ...669.1167B}, where the stellar age may be calculated using the color of the star and its rotation rate. As stars and their attendant planets form contemporaneously from the same clump of material, the stellar age can be taken as representative of the age of its planets. In addition, the relative alignment of the system may be learned by comparing stellar rotation periods determined from photometry to measurements of rotation from spectroscopic line broadening, or vsini, to calculate the stellar inclination. Although measuring the stellar inclination does not provide a measurement of the absolute angle between the planetary orbital plane, in the case of transiting planet systems any substantial inclination of the star implies misinclination between the stellar rotation and planetary orbit. Finally, the stellar rotation period may be compared with the planetary orbital periods to search for spin-orbit synchronization, hinting at the presence of tidal evolution in these systems.  

In this paper, we report on the rotation periods and variability properties  for $\sim$950 of the {\em Kepler} planet candidate host stars. In the following section, we describe the data, methods and metrics used to measure and quantify their variability and rotation periods. Section \ref{varprop} reports the general variabillity properties of the sample, rotation periods, and inferred ages for our sample, Section \ref{spinorbit} presents candidate systems with misalignment between the stellar rotation and planetary orbital axes, and Section \ref{tidal} discusses the possibility that several of the close-in planet candidates have been tidally synchronized to the spin of their host stars. We summarize our results and conclusions in Section \ref{conclusions}.

\section{DATA AND ANALYSIS}

We chose to work with the lightcurves from {\em Kepler} Quarter 9, as this was the first quarter  reduced using the PDC-MAP \citep{2010SPIE.7740E..62T} detrending pipeline (previous quarters have since been reprocessed, such that all {\em Kepler} data have now been detrended using PDC-MAP). PDC-MAP differs significantly from earlier iterations of the {\em Kepler} pipeline, in that it removes instrumental trends by fitting cotrending basis vectors that represent common non-astrophysical trends in the data. PDC-MAP is still subject to confusion when the timescale or morphology of astrophysical trends (i.e. stellar variability) are similar to that of instrumental effects (e.g. long term drifts caused by spacecraft motion), but uses knowledge of the ensemble of stars near the source of interest in the focal plane to constrain the coefficients used to fit the cotrending basis vectors. As such, PDC-MAP is quite robust, and leaves astrophysical variability intact the majority of the time. In this paper we work exclusively with the long cadence data, consisting of samples every 29.5 minutes.



We compute Lomb-Scargle periodograms \citep{1982ApJ...263..835S} and variability statistics for all stars in the {\em Kepler} exoplanet search sample \cite[please see][for a complete description of the variability metrics computed, and the properties of the {\em Kepler} exoplanet search targets as a whole]{2010ApJ...713L.155B,2011AJ....141...20B}. Here, we concentrate on finding rotation periods for the Kepler Objects of Interest (KOIs), or planet candidate host stars. Data that is as high quality and as nearly contiguous as the {\em Kepler} photometry can be both a blessing and a curse, as periodograms typically reveal a wealth of significant peaks. Some of these peaks are representative of true periodicities in the data, while others may be harmonics. Indeed, nature sometimes conspires to create stars with symmetric magnetic features on opposing hemispheres, such that the photometric period appears to be half of its true value \citep[e.g.][]{2009MNRAS.400..451C}. It also bears remembering that stars are not solid bodies, but rather rotate differentially, so the period found from a given set of observations represents the period(s) of the dominant feature(s) during the time of observation. There may therefore be multiple significant peaks at adjacent periods, broadening of individual peaks, or asymmetric lobes on individual peaks, depending on the presence and surface distribution of spots, as well as the latitudinal dependence and magnitude of differential rotation, and the presence of spot evolution \citep[see][for a more detailed discussion of the above issues]{2013ApJS..205...17W}. We therefore used our periodograms to cull a sample of periodic stars from the KOI host stars reported in \citet{2013ApJS..204...24B}, selecting only those lightcurves whose strongest peak had a power of at least 800, and period of less than 45 days (as the length of our dataset was a single quarter, or $\sim$90 days). Our simple threshold for what peak powers we deemed to be periodic was arrived at after extensive comparison between the periods found from our periodogram analysis and visual inspection of stars from the entire sample of $\sim$160,000 exoplanet search targets \citep[as described in][]{2011AJ....141...20B}. After using the results of the periodogram analysis to cull stars that fit these two criteria, we examined the lightcurves for our sample by-eye to confirm the periods found. In cases where the period found appeared to be a harmonic of the true period (i.e. when the period found appeared to be half the true period), we adjusted the period by folding the lightcurve over a range of trial periods within 5$\%$ of the next strongest peak in the periodogram and selecting the period that minimized the mean scatter in the folded lightcurve. It is of course still possible that nature may conspire to create perfectly symmetric spots on either hemisphere of the star, in which case the period found even with a check by-eye would still appear to be half of the true period, but we have made every effort to avoid these cases. 

\begin{figure}
\begin{center}
\includegraphics[width=0.5\textwidth]{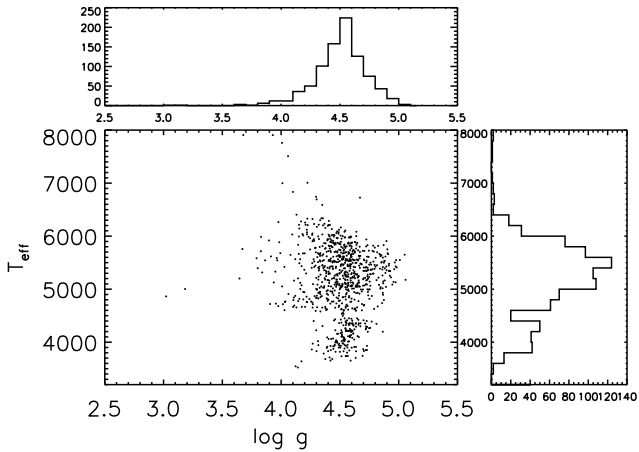}
\end{center}
\caption{Effective temperatures and gravities from the {\em Kepler} Input Catalogue for the sample of KOI host stars discussed in this paper. As with the majority of the {\em Kepler} targets, most of the stars selected here are FGK dwarfs, though there are some hotter stars and giants included as well. Panels above and to the right of the central panel show histograms of log(g) and effective temperature for the sample.}
\label{teffloggsample}
\end{figure}

The distribution in effective temperature and gravity for our selected sample is shown in Figure \ref{teffloggsample}. The sample is comprised largely of FGK dwarf stars (which make up the bulk demographic of the {\em Kepler} exoplanet search targets). However, as we carried out our sample selection by looking solely for highly periodic stars and did not apply any cuts on log(g) or effective temperature, some of the stars in the resulting sample have T$_{eff}$ $>$ 6200 K, and some are giants. We also note that the astrophysical sources of variability in our sample are therefore not solely due to stellar magnetism, and may also include pulsations, or geometrical variability as in the case of KOIs that are in fact eclipsing binaries rather than planets. As the majority of our sample are solar-like stars with outer convective envelopes, we expect rotational modulation by stellar magnetic features to be the dominant source of periodic variability for our stars (and indeed, it may appear in addition to some of the other astrophysical causes of variability, as with magnetically active eclipsing binaries that show both spot and geometric variability). In addition, it is also possible that the variability observed is associated with contamination from a background target or close companion that has fallen in the {\em Kepler} source aperture \citep[contamination has been reported for at least one object in the current sample, KOI-42;][]{2012ApJ...756...66H}. We do not expect this to be the case for the majority of targets, however, as targets that become KOIs to begin have typically been subjected to (and passed) a battery of tests that identify aperture contamination \citep[for further details see][]{2010ApJ...713L.103B}.

\section{RESULTS}

\subsection{Variability Properties, Rotation Periods and Inferred Ages} 
\label{varprop}

In the standard picture of stellar magnetic activity, faster rotation is associated with more vigorous dynamo action and correspondingly greater magnetic field. The stronger field results (through mechanisms that are as yet poorly understood) in heating of the outer atmosphere of the star, creating a hot chromosphere and even hotter corona. In addition, ropes of magnetic flux emerge from the stellar surface, whose footpoints create cool, dark starspots in the stellar photosphere as well as bright plage regions. The resulting observed relationship between stellar activity and rotation therefore manifests in a variety of tracers of stellar activity across the electromagnetic spectrum, tracing different conditions throughout the stellar atmosphere. 

{\em Kepler}'s broad optical bandpass photometric measures variability caused by starspots rotating into and out of visibility as the star spins. As such, it might be expected that more rapid rotation would lead to either an increased number of spots, and/or larger spots overall. One might additionally expect that the amplitude of photometric variability would share some relationship to complementary tracers of magnetic activity throughout the atmosphere, such as the commonly-used chromospheric activity index, log R$^{\prime}_{HK}$. Although these tracers are not exactly analogous (they originate in different parts of the stellar atmosphere, and log R$^{\prime}_{HK}$ also measures activity at a single point in time, whereas the amplitude of photometric variability is a single number that characterizes activity during the entire period of observations), they are both caused by the stellar magnetic field and therefore offer complementary information.

\begin{figure}
\begin{center}
\includegraphics[width=0.5\textwidth]{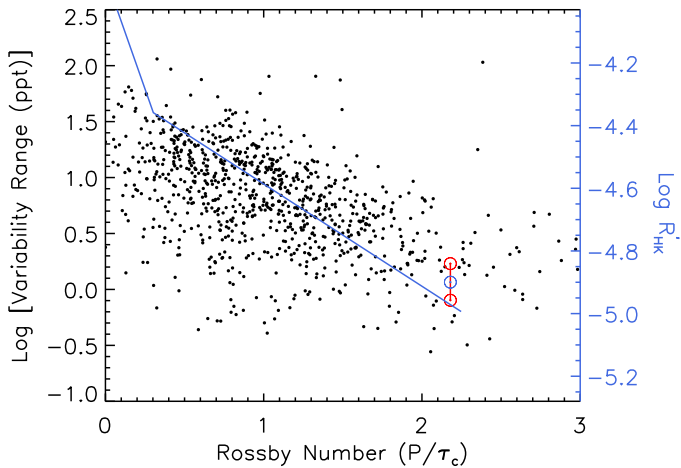}
\end{center}
\caption{The variability range, or amplitude of photometric variability, has a much larger spread of values for a given Rossby number than the commonly used Ca II spectroscopic activity index, log R$^\prime_{HK}$. Black points denote the Rossby number for {\em Kepler} Object of Interest host stars versus the log of the variability range in parts per thousand, or amplitude of the photometric variability, measured from {\em Kepler} Quarter 9 photometry. The blue line (and corresponding righthand axis) shows the relationship between the Rossby number and log R$^\prime_{HK}$, as reported by \citet{2008ApJ...687.1264M} for dwarfs in the solar neighborhood. The blue solar symbol denotes the Rossby number and  log R$^\prime_{HK}$ values for the Sun, adopted from \citet{2008ApJ...687.1264M}; the two red solar symbols, connected by a bar, show the spread of the solar photometric variability range as measured by SOHO (Basri, Walkowicz \& Reiners 2012). The axes are normalized such that the log R$^{\prime}_{HK}$ of the Sun falls in the middle of the range of observed photometric amplitudes. The spread in the solar values hints at the fact that even in an individual star, the variability range is dependent on the presence and particular surface distribution of spots at the time of observation, and may change over time. }
\label{rossbyrange}
\end{figure}

Figure \ref{rossbyrange} shows a comparison between photometric activity (as measured by the amplitude of stellar photometric variability during {\em Kepler} Quarter 9) versus the Rossby number\footnote{The Rossby number is a measure of whether the fluid dynamics in a system are dominated by rotation; in the context of stars, the Rossby number is the ratio of the stellar rotation period to the timescale of convective overturn \citep{1984ApJ...279..763N}. The timescale of convective overturn here is taken from \citet{1984ApJ...279..763N} Equation 4, based on the timescale at the base of the convection zone as determined from the models of \citep{1980LNP...114...19G}, and is a function only of spectral type.} for our sample of {\em Kepler} exoplanet candidate host stars. The overplotted blue line shows the relationship between chromospheric emission (as measured by log R$^{\prime}_{HK}$) and Rossby number, as determined by \citet{2008ApJ...687.1264M} for stars in the solar neighborhood; the blue solar symbol indicates the location of our Sun in log R$^{\prime}_{HK}$ and Rossby number, while the red solar symbols show the spread in the solar photometric variability between the active and quiet parts of the solar cycle \citep{2010ApJ...713L.155B} (the normalization between the two axes was set such that the log R$^{\prime}_{HK}$ of the Sun falls in the middle of the range of observed photometric amplitudes). For stars of a given Rossby number, there is clearly large scatter in the amplitude of photometric variability measured from the {\em Kepler} photometry, whereas the relationship between log R$^{\prime}_{HK}$ and Rossby number is considerably tighter. This scatter is likely caused in part by differences in the distribution of starspots over the stellar surface: if the variability is dominated by a monolithic spot or large group of spots on a relatively confined area of the stellar surface, the rotation of this feature into and out of view will create a higher amplitude of variability than spots that are more evenly distributed over the surface (such that the apparent area of spots viewed at any one time changes little, despite a large fraction of the surface being covered by magnetic features). Stellar optical photometric variability is therefore somewhat ambiguous as an indicator of the overall activity of the star-- while higher variability is likely due to an increased spotted area on the star, low amplitude variability may either be caused by a relatively unspotted star, or by a star with more evenly distributed spots. 

To further complicate matters, photometric variability comprises not only dark starspots but bright facular regions. It has been suggested that in some cases, the stellar surface may be dominated by bright network \citep[causing the otherwise featureless photosphere to appear relatively cool and therefore act as effective spots;][]{1992SoPh..142..197P}.
In spite of its ambiguities, however, photometric variability does contain information on the surface distribution of spots and active regions, and so provides not only a measure of the stellar rotation period but information on the geometry of the surface features of the field. Complementary measurements of the chromospheric activity of the star (e.g. from near-UV emission or log R$^{\prime}_{HK}$) provide a more definitive instantaneous measurement of the magnetic activity, and can be used to break the degeneracy between stars that have low amplitude variability caused by a few small spots, and those that have many spots with a surface distribution that leads to low amplitude variations. The synergy between spectroscopic and photometric measures of variability is particularly promising when trying to reconstruct the surface distribution of magnetic features using spot modeling, as the pristine, unspotted brightness of the star is typically unknown, and so features that do not create large amplitude variability (e.g. distributed spots, or spots near the pole that are always visible) would otherwise go undetected.

The stellar rotation periods determined for these planetary candidate host stars can also be used to constrain the age of the planetary system, through the application of gyrochronology relationships. As the star's age is typically a very difficult quantity to determine precisely (unless the star is located in a coeval population, has detectable lithium, or is amenable to detailed asterosesimic modeling), the application of gyrochronology relations is often the best opportunity to determine how old a given system is.   \citet{2007ApJ...669.1167B} developed a formalism for the relationship of stellar rotation and age, such that

\begin{eqnarray}
P(B-V,t) & = & f(B-V) \times g(t) \\
f(B-V) & = & a[(B-V)_0 - c]^b\\
g(t) & = &  t^n
\end{eqnarray}


These relationships state that the rotation period of the star (P) is a function of its color (B - V) and evolution over time (g(t)), relying on fit coefficients a, b, c, and n. Gyrochronology relationships are best calibrated for relatively young stars in clusters and associations, where isochronal fitting yields an independent determination of the age of the stellar population. The exact values for the fit coefficients a, b, c and n have been determined in a variety of contexts: \citet{2008ApJ...687.1264M} used a combination of the \citet{2007ApJ...669.1167B} relation, the Ca II activity index log R$^\prime_{HK}$, and the \citet{1984ApJ...279..763N}  relationship between age and Rossby number to determine ages for 108 F-K stars in the solar neighborhood out to the age of the Sun. The color dependencies for these relationships have also been calibrated using stellar clusters \citep[e.g. M35, M34, and Coma Ber][]{2009ApJ...695..679M,2011ApJ...733..115M,2009MNRAS.400..451C}. Colors were determined by transforming the SDSS filter photometry provided in the Kepler Input Catalog \citep{2011AJ....142..112B, 2007ASPC..364..165I}. It should be noted here that the gyrochronology relationships describe the spin-down of stars whose spin evolution has proceeded without significant external influences, for example tidal interaction from companion stars or close-in giant planets. If a star has an unidentified companion that has influenced its spin evolution, gyrochronology relations will yield an estimate that is younger or older than the true age of the star. Furthermore, the ages presented here assume that these stars lie on the so-called ``interface'' or ``I-sequence'' \citep[applying these relationships to ``C'' sequence stars will yield an underestimate of the stellar age, see][]{2008ApJ...687.1264M}.

\begin{figure}
\begin{center}
\includegraphics[width=0.5\textwidth]{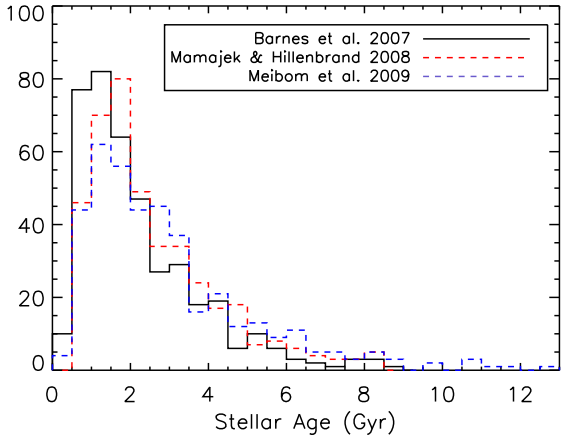}
\end{center}
\caption{The determination of rotation periods allows an estimate of the stellar age to be determined from gyrochronology relationships. As the rotation periods reported in this paper have all been determined from a single quarter of {\em Kepler} data, spanning 3 months, the rotaiton periods are typically on the order of a month or less. therefore, the distribution of the stellar ages tends to be young (the age of the Sun or less, with a few exceptions). Here we show histograms determined from three different gyrochronology relationships: Barnes et al 2007 (black solid line), Mamajek \& Hillenbrand 2008 (red dashed line) and Meibom et al. (2009) (blue dashed line). These three relationships produce comparable resulting age distributions; throughout we adopt ages predicted by the Mamajek \& Hillenbrand gyrochronology relations. }
\label{stellarage}
\end{figure}

In Figure \ref{stellarage}, we compare histograms of ages for the stars in our sample, determined using three gyrochronology relationships from the literature \citep{2007ApJ...669.1167B,2008ApJ...687.1264M,2009ApJ...695..679M}. We limit the sample of objects plotted here to the solar-type stars for which these relationships are best calibrated: those with color 0.5 $\ge$ B - V $\ge$ 0.9, T$_{eff}$ $\le$ 6200K, and P $\ge$ 10 days. The three gyrochronology relations used yield roughly comparable age distributions, with the only difference being a slight tendency for the \citet{2007ApJ...669.1167B} relationship to predict a relatively greater number of young stars in the sample. We adopt the gyrochronology coefficients determined by \citet{2008ApJ...687.1264M}, as their sample of solar-type nearby stars most closely resembles the {\em Kepler} exoplanet search target stars, and these coefficients better reproduce the age-rotation relationships for both young clusters and the Sun. In Table \ref{allstars} we provide age predictions for all stars in the sample with B - V $\ge$ 0.5 and T$_{eff}$ $\le$ 6200K, but caution that these age estimates may be unreliable for very rapidly rotating stars and cool stars with redder colors. 

Figure \ref{stellarage} shows the distribution of ages in the sample for which we are reporting periods, but it should not be taken as representative of the overall distribution of ages in the Kepler field. The ages we report are heavily biased towards young stars, as we required strong periodicity and periods less than 45 days, both of which select for more active, rapidly rotating stars. The ability to measure a rotation period at all requires that starspots be fairly large and remain stable for several rotation periods (i.e. that the spot evolution timescale is long compared to the rotation period), and additionally requires that starspots are located at a latitude where they periodically disappear from view as the star rotates. As little is known about the distribution of starspots on the surfaces of stars, spot evolution timescales, or the relative contribution of other magnetic features such as plage, and how these quantities change as a function of rotation period, it is difficult to quantify selection effects due to whether stars are amenable to having their rotation periods determined via this kind of analysis. We expect that as more rotation periods are determined for stars in the Kepler data set, the distribution of gyrochronological ages will grow to be more representative of the field population as a whole.   

It should also be noted that ages determined from these relationships are subject to considerable uncertainty for stars older than the Sun; at the present time, gyrochronology relationships are best calibrated for open clusters up to 1 Gyr in age \citep[e.g.][]{2011ApJ...733L...9M}, with the Sun being the only calibration point for older stars. However, one of the unique opportunities presented by {\em Kepler} is the ability to determine rotation periods for the older, less active stars that are typically amenable to astroseismic modeling (and therefore precise age determinations from asteroseismology), as {\em Kepler}'s precision makes it possible to measure very low levels of spot variability in these stars, while the duration of its observations allow slower rotation periods to be recovered. By using these older stars to calibrate gyrochronology relationships for stars older than 1 Gyr, we anticipate that it will be possible to determine ages for these and other exoplanet candidate systems with much greater precision in the future, but such additional calibration lies beyond the scope of the current work. 



\subsection{System Inclinations from Photometric Periods Combined with Spectroscopic vsini}
\label{spinorbit}

The photometric period of a star provides a straightforward way of determining the stellar rotation period. Whereas spectroscopic measurements of the rotational velocity provided by line broadening are subject to uncertainty in the stellar inclination, yielding $v sin i$ rather than the equatorial rotation period, photometry yields a rotation period whenever there are magnetic surface features that rotate into and out of view. These features are also subject to some ambiguity, as stars rotate differentially with latitude, and the signal that dominates the periodicity is that of the largest feature, located at some unknown latitude that is unlikely to be exactly the equator. However, differential rotation is a small effect (our own Sun differs only 20$\%$ between the equatorial and circumpolar rotation periods) relative to the potentially large ambiguity in $v sin i$. When combined with spectroscopic measurements of the rotation period and spectroscopically-determined stellar parameters, periods found from {\em Kepler} photometry can yield the stellar inclination via the relation

\begin{equation}
i = \arcsin(P_\star vsini/2\pi R_\star)
\end{equation}

where the stellar inclination is measuring along the line of sight of the observer, such that 90 degrees means that the stellar rotation axis is orthogonal to the line of sight, and 0 degree means that the star is seen pole-on. 

In recent years, a number of exoplanets have been shown to be misaligned with their hosts stars \citep[e.g.][]{2010ApJ...718..575W,2011ApJ...740L..10N,2011ApJS..197...14D,2012Natur.487..449S,2012ApJ...756...66H,2012ApJ...757...18A}, predominantly in systems with hot stars orbited by large, close-in planets. Alignment is typically measured using the Rossiter-McLaughlin effect, where a transiting planet causes deformation in the shape of the stellar spectral line profile as it passes in front of a rotating star \citep[thus alternately blocking the blue- or redshifted limb of the star as it passes across the face of the star during transit;][]{1924ApJ....60...15R,1924ApJ....60...22M,2000A&A...359L..13Q}. The Rossiter-McLaughlin effect is easier to measure in the case of hot stars, whose rapid rotation create broad spectral lines, and for large planets whose transits block a greater fraction of the starlight, creating a larger deformation in the line as the planet transits. The exact shape of the line deformation yields the sky-projected angle between the stellar rotation axis and the planetary orbital plane. Other measures of the spin-orbit alignment have been determined via modeling of starspots \citep[e.g.][]{2011ApJ...740L..10N,2012ApJ...756...66H,2012Natur.487..449S}, where asymmetries in the shape of the planetary transit caused by the planet occulting a starspot are combined with models of the out-of-transit variability to map the relative alignment of the star and the planet's transit chord. This method has been applied in the case of several stars that closely resemble the Sun. 

In the case of transiting planet candidates, such as those found by {\em Kepler}, the relative inclination of the system can also be determined for stars that have both photometric and spectroscopic periods. Here, the inclination measured is the inclination in the viewer's line-of-sight, rather than the in the plane of the sky as is the case for Rossiter-McLaughlin measurements. For both methods, measuring the inclination of the stellar rotation axis does not provide a measurement of the {\em absolute} angle between the planetary orbital plane and the star. However, in the case of inclinations inferred from comparing v sini with the photometric period, the fact that there are transiting planet candidates means that any system where the stellar inclination deviates significantly from 90$^\circ$ is misaligned to some extent \citep{2012ApJ...756...66H}. 

\begin{figure}
\begin{center} 
\includegraphics[width=0.5\textwidth]{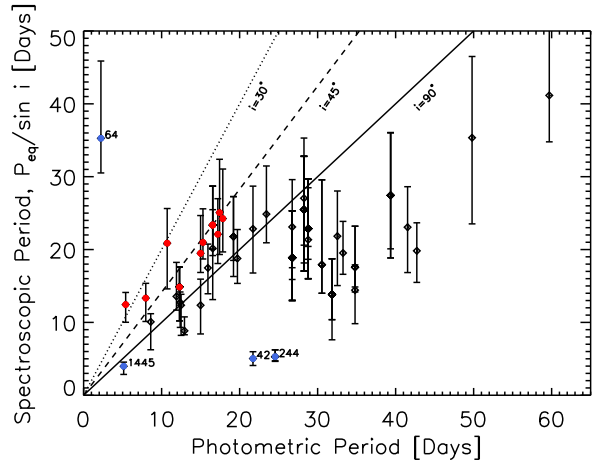}
\end{center}
\caption{Several of the KOIs show indications of stellar inclinations $<$90$^\circ$, implying misalignment between the axes of the stellar spin and planetary orbit. Here we show stellar rotation periods calculated from spectroscopic measurements of vsini and stellar parameters \citep{2012Natur.486..375B} versus rotation period determined from {\em Kepler} photometry (this work). The solid line indicates equality between the spectroscopic and photometric periods, or an equator-on inclination of 90$^\circ$, while the dashed and dotted lines indicate inclinations of 45$^\circ$ and 30$^\circ$, respectively. Beyond a rotation period of $\sim$25 days, the periods calculated from spectroscopy seem to flatten and fall below the rotation period calculated from the photometry. This effect may be due to the difficulty of measuring vsini for slower rotating stars, where vsini $<$ 2 km s$^{-1}$.  Red filled points show KOIs with possible spin-orbit misalignment. Points marked with blue filled points are KOIs 42, 64, 244, and 1445. KOI 42 has been noted by Hirano et al 2012 to have significant contamination from a background source, and KOI 64 has been determined to be a false positive. KOIs 244 and 1445 are relatively hot stars (T$_{eff}$ $>$ 6100K), where starspots are not expected to produce prominent spots, so the variability evident in their photometry may be related to either another astrophysical process or to contamination from background sources. }
\label{vsini}
\end{figure}

A small subset of our planet candidate host stars have published spectroscopic measurements of vsini and stellar parameters derived from high resolution spectra, gathered by the {\em Kepler} Follow-up Observing Program \citep[KFOP;][]{2010arXiv1001.0352G} using a wide variety of facilities; we refer the reader to \citet{2012Natur.486..375B} for the precise observations and methods used to extract the stellar parameters and rotational velocities. In Figure \ref{vsini}, we show the stellar rotation period calculated from measurements of vsini and the stellar radii reported in \citet{2012Natur.486..375B} versus the photometric rotation periods for these stars. Beyond periods of 25-30 days, the periods calculated from the \citet{2012Natur.486..375B} rotational velocities flatten out, becoming shorter than the periods determined from the {\em Kepler} photometry; this flattening is indicative of a loss of sensitivity to measurements of slow rotation periods, as the reported rotational velocities have ``error floor'' of 0.5 km s$^{-1}$ \citep{2012Natur.486..375B} and roughly similar radii (since the bulk of the {\em Kepler} sample is concentrated on solar analogues). However, for periods below $\sim$25 days, there is a clear relationship between the spectroscopic and photometric periods, including numerous targets where the spectroscopic period is longer than the photometric period (indicative of a non-90$^{\circ}$ stellar inclination). It is still possible that a symmetric spot distribution may cause one to measure half of the true rotation period, but we remind the reader that the periods reported have been checked by visual inspection of the lightcurve. Although a periodogram might be easily fooled by even a somewhat symmetric spot distribution on opposing hemispheres, the spots would have to be identical on both hemispheres to fool inspection by the human eye. Targets whose spectroscopic periods are consistent with sin i $<$ 1 are plotted in red; unfilled black diamonds show stars whose spectroscopic periods are consistent with sin i = 1; objects in blue have known periodicities that are not due to stellar rotation (see table notes for details). Overplotted are a solid line of equality, indicating 90$^{\circ}$ inclination, as well as lines showing the locus of 45$^{\circ}$ inclination (dashed) and 30$^{\circ}$ inclination (dotted). Inclinations, photometric periods, and periods inferred from spectroscopy for all systems are shown in Table \ref{spinorbittable}.

\citet{2010ApJ...718..575W}, \citet{2010ApJ...719..602S}, \citet{2012ApJ...757...18A} and \citet{2011ApJ...729..138M}, among others, have demonstrated that exoplanets around hot stars (T$_{eff}$ $>$ 6250K) are more likely to be misaligned, so we  examined whether the misaligned systems in our sample had preferentially hotter hosts. We performed a Komolgorov-Smirnov (K-S) test to compare the logg and effective temperature distributions of these misaligned systems to the rest of the sample. We found no evidence for a difference in effective temperature (K-S test yields D = 0.3 and P = 0.28), but there was mild evidence for a difference in logg between the two samples (D = 0.45 and P = 0.026, or a 2.2$\sigma$ difference). As the {\em Kepler} Input Catalog has documented systematic uncertainties in some stellar parameters \citep{2011AJ....142..112B}, we also checked against the stellar parameters hosted by the NASA Exoplanet Archive\footnotemark\footnotetext{exoplanetarchive.ipac.caltech.edu}, which incorporate stellar parameters derived from both adjustments to the original KIC parameters, as well as stellar parameters derived from follow-up spectroscopy (when available). Using the NASA Exoplanet Archive stellar parameters, we confirm that there is no evidence for an underlying difference in the temperature distribution for the misaligned systems (D = 0.3, P = 0.29), and find less evidence for a difference in the logg distributions (D = 0.36, P = 0.13). We conclude that the misaligned systems in our sample do not seem to preferentially occur around hot stars, although it is still possible that there are underlying systematic errors in measuring vsini for hot and cool stars (for example, hot stars tend to rotate more rapidly, making it easier to determine their rotation velocities compared to the more slowly rotating cool stars).

\subsubsection{The Curious Case of Kepler-9}

\begin{figure}
\begin{center}
\includegraphics[width=0.5\textwidth]{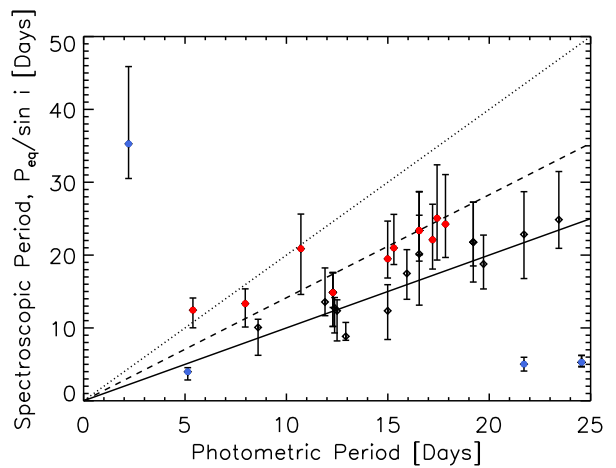}
\end{center}
\caption{Detail of Figure \protect\ref{vsini}, showing stellar rotation periods calculated from spectroscopic measurements of vsini and stellar parameters \citep{2012Natur.486..375B}  versus rotation period determined from {\em Kepler} photometry (this work) for object with P$_{phot}$ $<$ 25 days. }
\label{vsini_det}
\end{figure}

Almost all of the potentially misaligned systems in Figure \ref{vsini_det} are single planet systems, with one exception: the first multiple transiting planet system, Kepler-9 \citep{2010Sci...330...51H}, is marginally consistent with having an inclination of 45$^{\circ}$ ($\pm$ 10$^\circ$). The vsini of this system \citep[2.2 kms$^{-1}$, ][]{2012Natur.486..375B} is quite low, and is subject to fairly large uncertainty-- however, we point out that the true uncertainty in velocity is skewed, in the sense that while it would be difficult to measure a {\em lower} vsini, it would be relatively easy to measure if Kepler-9 were rotating more rapidly. Therefore, while it is possible that the vsini of Kepler-9 is lower than reported in \citet{2012Natur.486..375B}, it is unlikely to be rotating rapidly enough to make its spectroscopic period equal to its photometric period. 

We draw attention to this system as its three confirmed planets all transit the parent star, implying rough coplanarity \citep[although the three planets transit with somewhat different impact parameters-- Kepler-9b, c, and d have impact parameters of 0.35 $\pm$ 0.068, 0.621 $\pm$ 0.048, and 0.02 $\pm$ 0.22, respectively-- so this system is not as coplanar as many of the multiplanet systems found by {\em Kepler};][]{2012arXiv1202.6328F}. As previously stated, a measure of the stellar inclination does not yield the absolute angle between the stellar rotation axis and the orbital plane, but the significant misalignment of the star and the presence of transiting planets implies that some misalignment does exist. If this system is confirmed to be misaligned, it will be the first system in which three roughly coplanar planets exist in orbits misaligned from the rotation axis of their host star. 

If the Kepler-9 system is misaligned, it will be significant for a variety of reasons. Kepler-9 differs significantly from the hot stars hosting Jupiter-radius planets that dominate the known misaligned systems: the star is a G dwarf very similar to the Sun (T$_{eff}$ = 5722K, logg = 4.77), and the three transiting planets are relatively small (8.28R$_e$, 8.22R$_e$, and 1.67R$_e$).  A variety of mechanisms have been invoked to explain the orbital evolution and subsequent obliquities of hot Jupiters, including migration through the disk \citep{1996Natur.380..606L,2007A&A...473..329C}, Kozai oscillations combined with tidal friction \citep{2007ApJ...669.1298F}, planet-planet scattering\citep{2008ApJ...678..498N}, and misalignment between the stellar rotation axis and the protoplanetary disk \citep[e.g.][]{2010MNRAS.401.1505B,2011MNRAS.412.2790L,2013arXiv1304.5166B}. Most of the above scenarios predict that close-in giant planets eventually undergo realignment through tidal interaction with their parent star, provided the host star has an outer convective envelope to provide efficient tidal dissipation \citep[hence the observed dichotomy between obliquities of planets around hot and cool stellar hosts][]{2012ApJ...757...18A}. Tidal effects are not expected to be significant for smaller planets and planets in longer orbits \citep[and indeed the three known exoplanet systems that orbit stars with convective envelopes but still have high obliquity meet these criteria][]{2012ApJ...757...18A}. Planet-planet scattering can act on planets of any mass, but would not be expected to preserve coplanarity between the planets themselves. In the case of Kepler-9, the range of impact parameters for the three components intriguingly suggests that these planets are not as tightly aligned (within 2$^\circ$) as many other multiplanet systems \citep{2012arXiv1202.6328F}, and coincidental near-coplanarity after planet-planet scattering cannot be ruled out by the present work.

To misalign small planets while preserving coplanarity would likely require a primordial origin, wherein the disk became misaligned with the star early on and subsequently formed planets in misaligned orbits. Several mechanisms for misaligning a protoplanetary disk have been suggested in the literature, including gravitational perturbation by a companion or neighboring stars \citep[e.g][]{2010MNRAS.401.1505B,2013arXiv1304.5166B}, or interaction between the stellar magnetosphere and inner edge of the disk \citep{2012MNRAS.423..486L}. Reorientation of disks around young stars does seem to occur in nature, as observed S-symmetry bending in protostellar jets has been interpreted as the signature of changes in disk orientation over time (J. Bally \& R. Lovelace, private communication, but see also \citet{2009ApJ...692..943C} for observations in a massive star). If disk reorientation and misalignment occurs in nascent planetary systems, there should be a population of misaligned smaller planets in addition to the misaligned hot Jupiters orbiting hot stars. Further, as many of the mechanisms for reorienting the disk rely on external gravitational perturbation of the system, there should be some correlation between the occurrence of misaligned systems and binarity, environment, or both. A more complete census of the alignment of exoplanetary systems should provide interesting clues to planet formation scenarios.

\subsection{Possible Tidal Interaction in Close-In KOIs} 
\label{tidal}

\begin{figure}
\begin{center}
\includegraphics[width=0.5\textwidth]{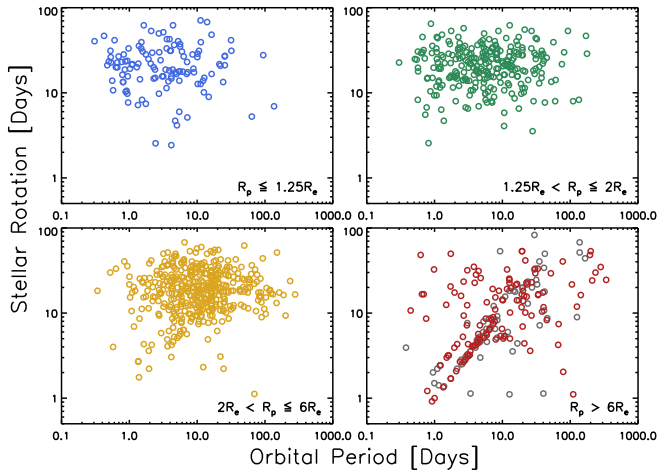}
\end{center}
\caption{For smaller planets, the stellar rotation period and planet orbital period are largely unrelated, but some KOIs with candidates above R$_p$ $\sim$ 6R$_\Earth$ show signs of possible interaction between the stellar rotation and planetary orbit. Here we show stellar rotation versus planetary orbital period, where symbol size and color denotes the size of the planetary candidate: blue circles, R$_p$ $\le$ 1.25R$_\Earth$; green circles, 1.25R$_\Earth$ $<$ R$_p$ $\le$ 2R$_\Earth$; gold circles, 2R$_\Earth$ $<$ R$_p$ $\le$ 6R$_\Earth$; red circles, 6R$_\Earth$ $<$ R$_p$ $\le$ 15R$_\Earth$; and gray circles, R$_p$ $>$ 15R$_\Earth$.}
\label{tidal_panels}
\end{figure}
\begin{figure}
\begin{center}
\includegraphics[width=0.5\textwidth]{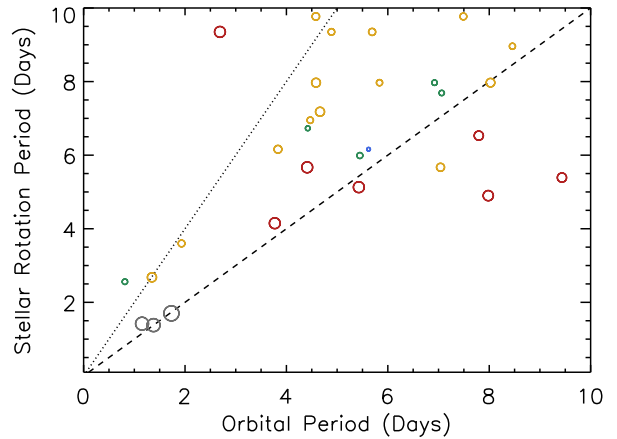}
\end{center}
\caption{Numerous candidates have orbital periods at or close to the stellar rotation period, implying dynamical interaction between the star and planet. Here the orbital period versus the stellar rotation period is shown for all KOIs with stellar rotation periods $<$ 10 days and planetary orbital periods $<$10 days, except those known to be false positives (see Table \protect\ref{sync}). The dashed line indicates exact synchronization, while the dotted line denotes the stellar rotation period being twice the planetary orbital period. All candidates lying along these two lines have radii of greater than 6R$_\Earth$, with the exception of KID 1725415 (KOI 2988). KOI 2988 has a radius of only 0.7R$_\Earth$, yet orbits almost exactly synchronously with its host. }
\label{tidalclean}
\end{figure}

We also carried out a search for signs of tidal interaction in our sample by comparing the rotation periods of the stars with the orbital periods of the planet candidates. In Figure \ref{tidal_panels}, we show the stellar rotation versus the planetary orbital period, binned by the radius of the planetary candidates (see figure caption for details). While there does not appear to be a relationship between stellar rotation and planetary orbital periods for planetary candidates smaller than R$_p$ = 6R$_e$, there is notable structure in the lower right panel for planetary candidates with  R$_p$ $>$ 6R$_e$, such that a number of these candidates have stellar rotation periods that are equal to that of the planetary orbital period. There are two reasons for this structure: first, a number of these planetary candidates have been found to be eclipsing binaries, where tidal synchronization has taken place. However, a small subset of these candidates have survived the vetting process thus far and remain viable planetary candidates. Figure \ref{tidalclean} shows the relationship between the stellar rotation and orbital periods for only candidates that have {\em not} been flagged as false positives; the dashed line in this plot indicates that the stellar rotation and planetary orbital periods are equal, while the dotted line indicates a stellar rotation period that is twice as long as the planetary orbital period.  A number of candidates lie on or very near these two loci, and there is a compelling dearth of candidates in these period ranges (P $<$ 6 days) that do {\em not} lie on these loci \citep[this paucity of planets was also recently noted by][]{2013arXiv1308.1845M}. A full list of candidates that lie within 10$\%$ of these loci are provided in Table \ref{tidaltable}, along with the computed ratios between the rotation and orbital periods. It is of course still possible that these candidates will be found to be false positives; in particular we note that many of them are very large radii (R$_p$ > 15 R$_e$), which have often turned out to be eclipsing binaries \citep[e.g.][]{2012A&A...545A..76S}, and in one case, a mutually eclipsing post-common envelope binary \citep{2013ApJ...767..111M}. However, the quality of the {\em Kepler} photometry is such that the overall false positive rate is quite low \citep[e.g.][]{2013ApJ...766...81F}, and as previously stated these candidates have survived the vetting process up until this point. Even if they are eclipsing binaries, the rotation period of the spotted star in these systems does match the orbital period of the companion. This fact lends credence to the two objects being physically associated, due to the very small chances of a spotted star being blended with an eclipsing binary that had exactly the same orbital period as its rotation period. Below, we consider the implications if these Objects of Interest are in fact confirmed to be true planets.
 
It has long been known that the components of stellar binary systems exchange angular momentum, causing their orbital and rotation periods to synchronize and circularize over time; this interaction leads to observable differences in the magnetic activity of the system members than would be expected for solitary stars. In star-planet systems, tidal synchronization of the stellar rotation presents a puzzle: the exchange of angular momentum required to for a planet to spin up and become synchronized with its parent star would cause the planet's orbit to rapidly decay, making the planet spiral into the star before any significant spin-up can occur \citep{2010ApJ...718..575W}. However, the details of tidal theory are still highly uncertain \citep[in particular as regards the tidal quality factor Q, which is expected to vary by orders of magnitude for bodies of different masses in different orbital configurations;][]{2011ApJ...731...67P}. 

Several exoplanet systems reportedly show signs of spin-orbit coupling, including synchronization: \citet{1997ApJ...474L.119B} found that $\tau$ Boo displayed periodic variability in Ca II emission at the orbital period of its planet, and \citet{2009MNRAS.396.1789P} have presented empirical evidence that host stars of massive close-in planets seem to rotate more rapidly than stars without such planets\footnotemark\footnotetext{\citet{2009MNRAS.396.1789P} in fact specifically predict the existence of super-Jupiters orbiting near one day periods in sync with their host stars, akin to the systems in Figure \protect\ref{tidalclean}.}. In the case of $\tau$ Boo, \citet{1998AJ....115.2122D} posited that perhaps only the outer convective envelope of the star had been spun up, rather than the star itself (an appealing solution, as $\tau$ Boo is a late F dwarf, and so has a relatively thin convective envelope compared to less massive stars like the Sun). Alternatively, \citet{2009MNRAS.396.1789P} suggested that tidal interaction may partially compensate for the drag of the star's magnetized wind, causing the star to spin down more slowly than it would otherwise \citep[see also][]{2010ApJ...723L..64C}. The latter explanation would act over a wider range of stellar host masses, and could account for apparent spin-orbit coupling around all stellar hosts. We tested whether the stellar hosts in systems near the locus of synchronicity (i.e. those with period ratios between 0.9 and 1.1) were preferentially around hotter stars, but found no evidence that this was the case (a K-S test of the effective temperature and logg distributions yielded D$_{logg}$ = 0.24, P$_{logg}$ = 0.51 and D$_{T_{eff}}$ = 0.21, P$_{T_{eff}}$ = 0.67, respectively). 

As our stellar rotation periods are measured from the periodic variability caused by surface magnetic features, we also considered the possibility that the period we measure could be due to an an induced magnetic feature on the star created by interaction with the planet, rather than the true rotation period of the star. Such star-planet interaction (SPI) has been detected in periodic variability of Ca II emission, thought to be due to enhanced emission on the stellar surface in the presence of magnetic interaction with a close-in planet \citep[see for example][though note that \citet{1997ApJ...474L.119B} essentially found a similar signature, but interpreted it as representative of the stellar rotation and therefore due to tidal synchronization]{2005JRASC..99...23S}. Searches for SPI have also uncovered correlations between the presence of planets and enhanced stellar activity, where in many cases it is not clear whether the observed activity enhancements are due to stellar spin-up (or stymied spin-down) by tidal interactions \citep{2013ApJ...766....9S}. If the variability we observe was due to a planet-induced magnetic bright or dark spot on the surface of the star, we would expect that the phase of the spot variability would be related to the planetary transit (i.e. the out-of-transit lightcurve would be at either its brightest or darkest during the transit). We find no evidence of this effect in a by-eye examination of the lightcurves for these candidates; the spot pattern does modulate the lightcurve at the orbital period of the planet, but in all cases the spot morphology evolves over the course of the lightcurve such that transits do not always occur at a preferred point in the spot pattern. The evolution of the out-of-transit lightcurve morphology is consistent with typical starspot modulation for solar-like stars. Of course, these observations do not exclude the possibility that star-planet interaction does exist in these systems, only that its signature is not readily discernible from stellar photospheric magnetic features; it may be possible that such a signature would only be evident in monitoring of chromospheric features (such as the Ca II H and K lines).

Finally, we point out an additional curious feature of Figure \ref{tidalclean}: two of the candidate planets where the stellar rotation period is twice the planetary orbital period are not the super-Jupiter or Jupiter radius planets for which tidal interactions are expected to be most effective, they are R$_p$ = 6R$_e$ and 2.64R$_e$. We can only speculate as to how these particular planets came to be in their current orbits-- it is intriguing to think that they might have once possessed larger gas envelopes that were subsequently eroded by the influence of stellar activity, or perhaps they were shepherded into their current positions through dynamical interactions elsewhere in the system. Certainly if these and the other, larger-radius candidates become confirmed planets, they comprise an interesting set of systems through which detailed modeling may lead to a better understanding of tidal effects in planetary system evolution.

\section{CONCLUSIONS AND FUTURE WORK}
\label{conclusions}

In this work, we have reported rotation periods, variability properties and gyrochronological ages for $\sim$950 of the {\em Kepler} planet candidate host stars. We find that for a given Rossby number, there is a wide range in the amplitude of the variability due to starspots. We interpret this scatter in the photometric amplitude as being caused by differing distributions of spots over the surfaces of the stars (which sometimes create relatively low amplitude variability even for stars with rapid rotation and accordingly high magnetic activity). We anticipate that a future direct comparison between the photometric variability measured by {\em Kepler} and spectroscopic indicators of chromospheric activity (such as the Ca II H and K lines) will provide additional clues to activity in these stars. We intend to pursue further analysis of additional quarters of data to obtain a complete set of rotation periods for the rest of the planet candidate hosts, as the long duration of the {\em Kepler} observations provides plenty of time for spots on a given star to evolve and create variability more amenable to a straightforward period determination. Models of the spot distributions on these and other {\em Kepler} target stars will additionally provide a sense of the latitudinal differential rotation, starspot coverage, and spot evolution timescales. 

We also compared the rotation periods found for our sample of KOIs with the planetary system properties. By comparing the rotation periods measured from the {\em Kepler} photometry with spectroscopic measurements of vsini and spectroscopically-derived stellar parameters, we identify a subset of systems that may have spin-orbit misalignment. One of these systems is Kepler-9, the first known exoplanet system with multiple transiting planets. We discussed a number of scenarios that have been invoked to explain spin-orbit misalignment, and argue that the mutual coplanarity of this triply-transiting system suggests a primordial origin where the current planets formed from a protoplanetary disk that became misaligned with the stellar spin early on. In addition, we find a small population of stars whose rotation periods seem to be either equal to or twice as long as the orbital periods of their planet candidates. In most cases these planet candidates have radii larger than Jupiter, although two such candidates have relatively small radii in the Neptune-radius regime (2 - 6R$_e$). We believe that the most likely explanation for these systems is some kind of tidal interaction between the planet and star, as the morphology of the starspot variability does not seem to be linked to the phase of the planetary transit. 

During the completion of this manuscript, the {\em Kepler} spacecraft suffered a failure of one of its three remaining reaction wheels, thus losing its ability to perform the fine pointing that plays an integral role in the collection of its precision lightcurves. As of this writing, it remains unclear whether {\em Kepler} will return to collecting data, and seems unlikely that it will return to collecting data for the exoplanet discovery mission that was its central goal. Although the untimely end of the mission is certainly a loss for both the exoplanet and stellar astrophysics communities, {\em Kepler} leaves behind a treasure trove of data that will be its legacy for many years to come. {\em Kepler} is dead; long live {\em Kepler}!

\acknowledgements
LMW and GB thank the entire {\em Kepler} team for their past and continued efforts in making the mission a great success. LMW thanks Kaloyan Penev, Jeff Oishi, Dong Lai, Jamie Lloyd, John Johnson, Doug Lin, Richard Lovelace and John Bally for illuminating conversations that contributed greatly to the writing of this manuscript. This paper includes data collected by the Kepler mission. Funding for the Kepler mission is provided by the NASA Science Mission directorate. This research has made use of the NASA Exoplanet Archive, which is operated by the California Institute of Technology, under contract with the National Aeronautics and Space Administration under the Exoplanet Exploration Program. All of the data presented in this paper were obtained from the Mikulski Archive for Space Telescopes (MAST). STScI is operated by the Association of Universities for Research in Astronomy, Inc., under NASA contract NAS5-26555. Support for MAST for non-HST data is provided by the NASA Office of Space Science via grant NNX09AF08G and by other grants and contracts.

\begin{deluxetable*}{lcccc}
\tablecaption{Gyrochronology Relation Coefficients}
\tablenum{1}
\tabletypesize{\footnotesize}
\tablehead{\colhead{Source} & \colhead{a} & \colhead{b} & \colhead{c} & \colhead{n} \\ 
\colhead{} &\colhead{} & \colhead{} & \colhead{} & \colhead{} } 
\startdata
\citet{2007ApJ...669.1167B} & 0.7725 $\pm$ 0.011 & 0.6010 $\pm$ 0.024 & 0.4000 $\pm$ 0.000 & 0.5189 $\pm$ 0.007 \\
\citet{2008ApJ...687.1264M}& 0.4070 $\pm$ 0.021& 0.3250 $\pm$ 0.024 & 0.4950 $\pm$ 0.010 & 0.5660  $\pm$ 0.008\\
\citet{2009ApJ...695..679M} & 0.7700 $\pm$ \nodata & 0.4720 $\pm$ \nodata & 0.5530 $\pm$ \nodata & 0.5200 $\pm$ \nodata \\
\enddata
\end{deluxetable*}

\begin{deluxetable*}{cccccccccc}

\tabletypesize{\footnotesize}

\tablecaption{Periods and Variability for Kepler Objects of Interest}
\tablenum{2}
\label{allstars}
\tablehead{\colhead{Kepler ID} & \colhead{KOI} & \colhead{T$_{eff}$} & \colhead{Logg} & \colhead{Log(Range)} & \colhead{Period} & \colhead{P$_{err}$} & \colhead{Ro} & \colhead{B - V} & \colhead{Age} \\ 
\colhead{} & \colhead{} & \colhead{(K)} & \colhead{(cm s$^{-2}$)} & \colhead{(ppt)} & \colhead{(days)} & \colhead{(days)} & \colhead{} & \colhead{} & \colhead{(Gyr)} } 

\startdata
5903312 & 8 & 5783 & 4.29 & 0.43 & 13.88 & 3.28 & 1.12 & 0.66 & 1.46 \\
7684873 & 14 & 7906 & 3.93 & -0.07 & 5.83 & 7.32 & \nodata & 0.24 & \nodata \\
7255336 & 19 & \nodata & \nodata & 0.15 & 2.43 & 2.40 & 0.13 & 0.78 & 0.05 \\
10125352 & 21 & 6122 & 4.22 & 0.29 & 4.59 & 1.03 & 0.67 & 0.56 & 0.35 \\
9071386 & 23 & 6324 & 4.27 & 0.32 & 4.69 & 9.07 & 0.93 & 0.52 & 0.62 \\
4743513 & 24 & 5890 & 4.42 & 0.47 & 2.08 & 7.12 & 0.24 & 0.59 & 0.07 \\
10593759 & 25 & 5995 & 4.32 & 0.98 & 3.13 & 0.52 & 0.31 & 0.62 & 0.12 \\
8866102 & 42 & 6035 & 4.22 & 0.03 & 21.71 & 6.60 & 3.85 & 0.53 & 7.26 \\
\enddata

\tablecomments{Full machine-readable table available online. Period uncertainties provided are taken as the half-width half-maximum of a gaussian with the same peak height and integrated area as the periodogram peak nearest the adopted period. These widths are largely a function of the periodogram sampling (longer periods tend to have wider peaks) and so are overestimates of the uncertainty in the period, but are provided here for completeness. For hotter stars (T$_{eff}$ $>$ 6200K) that are not expected to possess a solar-like magnetic dynamo, periodicities may be due to ellipsoidal variations or pulsations rather than rotational modulation by stellar magnetic features. Gyrochronological ages are provided for all objects having B$-$V $>$ 0.5 and T$_{eff}$ $<$ 6200K, but the reader is advised that these ages are highly uncertain for very young stars (P $<$ 10 days) and cool stars. }
\end{deluxetable*}

\newpage
\LongTables
\begin{deluxetable*}{rrrrrrc}
\tabletypesize{\footnotesize}
\tablecaption{Comparison of Stellar Rotation Period to Planetary Orbital Period for Planetary Candidates with P$_{orb}$ $<$ 10 days}
\tablenum{3}
\tablehead{\colhead{KID} & \colhead{KOI} & \colhead{P$_\star$} & \colhead{P$_{orb}$} & \colhead{P$_\star$/P$_{orb}$} & \colhead{R$_p$} & \colhead{Notes} \\ 
\colhead{} & \colhead{} &  \colhead{(days)} & \colhead{(days)} & \colhead{} & \colhead{(Earth radii)} & \colhead{} } 
\startdata
11554435 & 63 & 5.39 & 9.43 & 0.57 & 6.31 & \nodata \\
 6305192 & 219 & 7.97 & 8.03 & 0.99 & 5.00 & \nodata \\
 9139084 & 323 & 7.97 & 5.84 & 1.37 & 2.17 & \nodata \\
 9967884 & 425 & 5.13 & 5.43 & 0.95 & 11.30 & \nodata \\
 10973664 & 601 & 5.39 & 5.40 & 1.00 & 3.00 & \nodata \\
 7447200 & 676 & 12.2 & 7.97 & 1.54 & 3.30 & \nodata \\
 9963524 & 720 & 9.35 & 5.69 & 1.64 & 2.96 & \nodata \\
 6392727 & 851 & 7.97 & 4.58 & 1.74 & 5.40 & \nodata \\
 6948054 & 869 & 9.77 & 7.49 & 1.30 & 2.70 & \nodata \\
 7380537 & 883 & 9.35 & 2.69 & 3.48 & 11.12 & \nodata \\
 7767559 & 895 & 5.67 & 4.41 & 1.29 & 11.40 & \nodata \\
 7870390 & 898 & 11.54 & 9.77 & 1.18 & 2.83 & \nodata \\
 9480189 & 941 & 10.5 & 6.58 & 1.60 & 3.40 & \nodata \\
 10272640 & 1074 & 4.15 & 3.77 & 1.10 & 11.10 & \nodata \\
 8958035 & 1391 & 4.90 & 7.98 & 0.61 & 8.80 & \nodata \\
 7449844 & 1452 & 1.42 & 1.15 & 1.23 & 23.00 & \nodata \\
 9909735 & 1779 & 7.18 & 4.66 & 1.54 & 5.80 & \nodata \\
 11551692 & 1781 & 11.03 & 7.83 & 1.41 & 3.76 & \nodata \\
 6058614 & 1799 & 1.70 & 1.73 & 0.98 & 49.00 & \nodata \\
 11017901 & 1800 & 6.53 & 7.79 & 0.84 & 6.20 & \nodata \\
 11853878 & 1833 & 10.7 & 3.69 & 2.90 & 1.65 & \nodata \\
 9471268 & 1835 & 9.77 & 4.58 & 2.13 & 3.38 & \nodata \\
 7765528 & 1840 & 5.67 & 7.04 & 0.81 & 4.10 & \nodata \\
 10464050 & 1851 & 6.95 & 4.47 & 1.55 & 2.30 & \nodata \\
 4263293 & 1895 & 8.96 & 8.46 & 1.06 & 2.10 & \nodata \\
 6862721 & 1982 & 9.35 & 4.89 & 1.91 & 2.50 & \nodata \\
 6665512 & 2005 & 7.97 & 6.92 & 1.15 & 1.67 & \nodata \\
 9790806 & 2035 & 3.60 & 1.93 & 1.86 & 2.64 & \nodata \\
 10329835 & 2058 & 11.19 & 1.52 & 7.34 & 1.11 & \nodata \\
 6921944 & 2114 & 6.73 & 4.42 & 1.52 & 1.33 & \nodata \\
 8261920 & 2174 & 11.54 & 6.69 & 1.72 & 1.35 & \nodata \\
 6200235 & 2350 & 10.56 & 1.08 & 9.79 & 1.60 & \nodata \\
 8256453 & 2573 & 2.68 & 1.35 & 1.99 & 6.00 & \nodata \\
 5794570 & 2675 & 5.99 & 5.45 & 1.10 & 2.00 & \nodata \\
 6779260 & 2678 & 6.16 & 3.83 & 1.61 & 3.80 & \nodata \\
 8639908 & 2700 & 10.56 & 0.91 & 11.60 & 1.26 & \nodata \\
 5175986 & 2708 & 10.42 & 0.87 & 12.00 & 1.69 & \nodata \\
 5953297 & 2733 & 6.16 & 5.62 & 1.10 & 0.90 & \nodata \\
 7659389 & 2734 & 10.87 & 3.83 & 2.84 & 2.90 & \nodata \\
 11456382 & 2771 & 2.56 & 0.81 & 3.14 & 1.68 & \nodata \\
 7428736 & 2827 & 7.69 & 7.06 & 1.09 & 1.60 & \nodata \\
 6129524 & 2886 & 10.27 & 0.88 & 11.65 & 2.00 & \nodata \\
\hline
 7377033 & 882 & 3.92 & 1.96 & 2.00 & 18.00 & FP?\tablenotemark{a} \\
 7255336 & 19 & 2.43 & 1.20 & 2.02 & 13.00 & FP \\
 4743513 & 24 & 2.08 & 2.09 & 1.00 & 12.20 & FP \\
 10593759 & 25 & 3.13 & 3.13 & 1.00 & 26.00 & FP \\
 2445975 & 53 & 1.13 & 3.39 & 0.33 & 56.99 & FP \\
 8248939 & 61 & 4.32 & 1.63 & 2.64 & 2.49 & FP \\
 7051180 & 64 & 2.22 & 1.95 & 1.14 & 5.12 & FP \\
 11673674 & 133 & 4.79 & 4.62 & 1.04 & 7.60 & FP \\
 10904857 & 194 & 4.32 & 3.12 & 1.38 & 14.20 & FP \\
11548140 & 256 & 1.38 & 1.38 & 1.00 & 24.00 & FP \\
 11189127 & 347 & 2.68 & 2.67 & 1.00 & 49.00 & FP \\
 3230578 & 381 & 6.34 & 6.34 & 1.00 & 6.56 & FP \\
 9597411 & 424 & 3.18 & 1.58 & 2.02 & 64.00 & FP \\
 5122112 & 552 & 3.27 & 3.06 & 1.07 & 11.00 & FP \\
 5443837 & 554 & 4.07 & 3.66 & 1.11 & 7.70 & FP \\
 5608566 & 609 & 5.02 & 4.40 & 1.14 & 14.90 & FP \\
 10384962 & 619 & 3.09 & 2.88 & 1.07 & 22.30 & FP \\
 8409588 & 690 & 2.72 & 1.36 & 2.00 & 5.30 & FP \\
 8908102 & 699 & 5.41 & 5.41 & 1.00 & 141.00 & FP \\
 9162741 & 703 & 4.69 & 1.37 & 3.43 & 1.33 & FP \\
 9834719 & 715 & 3.18 & 1.62 & 1.96 & 56.00 & FP \\
 10068383 & 725 & 8.96 & 7.30 & 1.23 & 13.30 & FP \\
 10157573 & 726 & 5.13 & 5.12 & 1.00 & 3.40 & FP \\
 7270230 & 876 & 8.27 & 7.00 & 1.18 & 13.00 & FP \\
 3233043 & 966 & 3.92 & 0.38 & 10. & 56.00 & FP \\
 2157247 & 997 & 5.68 & 5.69 & 1.00 & 6.50 & FP \\
 5899544 & 1034 & 1.70 & 1.74 & 0.98 & 55.00 & FP \\
 5817553 & 1040 & 4.23 & 4.21 & 1.01 & 3.80 & FP \\
 8242681 & 1065 & 3.92 & 4.02 & 0.97 & 17.10 & FP \\
 10232123 & 1075 & 1.34 & 1.34 & 1.00 & 9.13 & FP \\
 8279765 & 1130 & 2.76 & 2.76 & 1.00 & 20.70 & FP \\
 10287248 & 1152 & 3.00 & 4.72 & 0.64 & 19.05 & FP \\
 3547091 & 1177 & 3.22 & 3.31 & 0.97 & 48.00 & FP \\
 6629332 & 1227 & 4.32 & 2.16 & 2.00 & 28.00 & FP \\
6387450 & 1228 & 4.07 & 3.66 & 1.11 & 13.10 & FP \\
 8488878 & 1248 & 5.83 & 5.80 & 1.00 & 11.60 & FP \\
 8620565 & 1250 & 1.21 & 0.78 & 1.55 & 15.80 & FP \\
 7199774 & 1346 & 4.79 & 4.71 & 1.02 & 30.00 & FP \\
 6866228 & 1348 & 7.97 & 7.70 & 1.03 & 20.00 & FP \\
 7220322 & 1350 & 8.60 & 0.75 & 11.43 & 8.58 & FP \\
 9451127 & 1381 & 5.13 & 5.12 & 1.00 & 27.00 & FP \\
 9446824 & 1382 & 4.79 & 4.20 & 1.14 & 23.00 & FP \\
 8953257 & 1383 & 3.18 & 3.22 & 0.99 & 28.00 & FP \\
 9002237 & 1389 & 4.40 & 4.35 & 1.01 & 20.20 & FP \\
 11517719 & 1416 & 2.50 & 2.50 & 1.00 & 16.30 & FP \\
 11599264 & 1417 & 8.27 & 8.43 & 0.98 & 14.60 & FP \\
 11100657 & 1418 & 5.13 & 8.34 & 0.61 & 11.90 & FP \\
 12506351 & 1446 & 2.43 & 1.23 & 1.98 & 76.00 & FP \\
 9705459 & 1448 & 2.83 & 2.49 & 1.14 & 21.00 & FP \\
 7532973 & 1450 & 2.22 & 2.14 & 1.04 & 85.00 & FP \\
 8081482 & 1539 & 2.80 & 2.82 & 0.99 & 30.00 & FP \\
 5649956 & 1540 & 2.43 & 1.21 & 2.01 & 36.00 & FP \\
 5270698 & 1543 & 4.15 & 3.96 & 1.05 & 15.50 & FP \\
 5475431 & 1546 & 0.92 & 0.92 & 1.00 & 9.50 & FP \\
 9940565 & 1548 & 4.23 & 2.14 & 1.98 & 26.00 & FP \\
 11043136 & 1644 & 5.26 & 2.62 & 2.01 & 9.200 & FP \\
 11546211 & 1654 & 2.22 & 1.10 & 2.02 & 11.55 & FP \\
 4832197 & 1661 & 2.01 & 0.95 & 2.12 & 95.00 & FP \\
 8043714 & 1765 & 8.60 & 6.81 & 1.26 & 14.10 & FP \\
 6153672 & 1794 & 3.54 & 3.36 & 1.05 & 41.00 & FP \\
 6507427 & 2455 & 4.69 & 4.74 & 0.99 & 2.40 & FP \\
 10747445 & 2673 & 5.26 & 4.59 & 1.15 & 23.80 & FP \\
 5475641 & 2895 & 9.35 & 1.07 & 8.74 & 1.60 & FP \\
 3228945 & 2917 & 5.53 & 0.73 & 7.57 & 1.78 & FP \\
 8192911 & 3058 & 8.96 & 7.54 & 1.19 & 2.16 & FP \\
  \enddata
\tablenotetext{a}{Identified as a possible eclipsing binary by \citet{2013arXiv1308.1845M}}
\tablecomments{FP denotes known false positives; candidates with that are not marked FP are plotted in Figure \ref{tidalclean}. False positives are as listed on the NASA Exoplanet Archive, \url[http://exoplanetarchive.ipac.caltech.edu/cgi-bin/ExoTables/nph-exotbls?dataset=cumulative]{http://exoplanetarchive.ipac.caltech.edu/cgi-bin/ExoTables/nph-exotbls?dataset=cumulative}, current as of January 3, 2013. A subset of these are known eclipsing binaries found in the current release version of the {\em Kepler} Eclipsing Binary Catalog, Revision 1.96, available at \url[http://keplerebs.villanova.edu]{http://keplerebs.villanova.edu}.}
\label{tidaltable}
\end{deluxetable*}

\clearpage

\begin{deluxetable*}{rrrrrrrrrrrr}
\tabletypesize{\footnotesize}
\tablecaption{Comparison of Stellar Rotation Periods Determined from Photometry and Spectroscopy}
\tablenum{4}

\tablehead{\colhead{KID} & \colhead{KOI} & \colhead{P$_{phot}$} & \colhead{P$_{spec}$} & \multicolumn{2}{c}{err P$_{spec}$} & \colhead{vsini} & \colhead{vsini err} & \colhead{R$_{\star}$} & \multicolumn{2}{c}{err R$_{star}$} & \colhead{inclination} \\ 
\colhead{} & \colhead{} & \colhead{(days)} & \colhead{(days)} & \multicolumn{2}{c}{(days)} & \colhead{(km s$^{-1}$)} & \colhead{(km s$^{-1}$)} & \colhead{(R$_{Sun}$)} & \multicolumn{2}{c}{R$_{Sun}$} & \colhead{(degrees)} } 

\startdata
8866102 & 42\tablenotemark{a} & 21.71 & 5.03 & +0.94 & -0.58 & 14.90 & $\pm$0.50 & 1.48 & +0.27 & -0.16 & \nodata \\
11554435 & 63 & 5.39 & 12.44 & +1.66 & -1.68 & 3.80 & $\pm$0.50 & 0.94 & +0.02 & -0.03 & 25.7 \\
7051180 & 64 & 2.22 & 35.27 & +10.60 & -8.17 & 2.30 & $\pm$0.50 & 1.60 & +0.33 & -0.13 & 3.6 \\
6850504 & 70 & 28.25 & 25.45 & +7.34 & -8.41 & 1.80 & $\pm$0.50 & 0.91 & +0.07 & -0.16 & 90. \\
7199397 & 75 & 21.71 & 22.84 & +5.86 & -3.60 & 5.50 & $\pm$0.50 & 2.48 & +0.60 & -0.32 & 90. \\
10187017 & 82 & 26.74 & 18.86 & +6.46 & -5.86 & 1.70 & $\pm$0.50 & 0.63 & +0.11 & -0.06 & \nodata \\
2571238 & 84 & 32.54 & 21.83 & +6.20 & -6.18 & 1.80 & $\pm$0.50 & 0.78 & +0.05 & -0.04 & \nodata \\
8505215 & 99 & 28.79 & 21.34 & +7.14 & -7.14 & 1.50 & $\pm$0.50 & 0.63 & +0.02 & -0.02 & \nodata \\
8456679 & 102 & 49.80 & 35.35 & +11.15 & -10.63 & 1.80 & $\pm$0.50 & 1.26 & +0.19 & -0.14 & \nodata \\
2444412 & 103 & 10.71 & 20.87 & +4.76 & -6.66 & 2.20 & $\pm$0.50 & 0.91 & +0.01 & -0.20 & 30.9 \\
8349582 & 122 & 59.70 & 41.15 & +12.98 & -11.63 & 1.90 & $\pm$0.50 & 1.55 & +0.27 & -0.16 & \nodata \\
9818381 & 135 & 12.38 & 12.74 & +1.43 & -1.42 & 4.60 & $\pm$0.50 & 1.16 & +0.03 & -0.03 & 90. \\
5735762 & 148 & 39.36 & 27.44 & +8.61 & -8.59 & 1.60 & $\pm$0.50 & 0.87 & +0.02 & -0.02 & \nodata \\
10925104 & 156 & 31.85 & 13.83 & +4.90 & -3.46 & 2.00 & $\pm$0.50 & 0.55 & +0.14 & -0.01 & \nodata \\
5084942 & 161 & 26.74 & 23.10 & +6.50 & -6.51 & 1.80 & $\pm$0.50 & 0.82 & +0.04 & -0.04 & 90. \\
9573539 & 180 & 15.94 & 17.48 & +3.27 & -4.18 & 2.70 & $\pm$0.50 & 0.93 & +0.02 & -0.14 & 90. \\
10619192 & 203\tablenotemark{b} & 12.50 & 12.36 & +1.50 & -1.50 & 4.20 & $\pm$0.50 & 1.03 & +0.03 & -0.02 & 90. \\
4349452 & 244\tablenotemark{c} & 24.55 & 5.29 & +0.94 & -0.56 & 10.90 & $\pm$0.50 & 1.14 & +0.19 & -0.11 & \nodata \\
8478994 & 245 & 28.79 & 22.85 & +6.86 & -6.85 & 1.70 & $\pm$0.50 & 0.77 & +0.05 & -0.05 & 90. \\
5383248 & 261\tablenotemark{d} & 15.30 & 20.98 & +4.61 & -6.05 & 2.30 & $\pm$0.50 & 0.95 & +0.03 & -0.18 & 46.8 \\
5088536 & 282 & 34.80 & 17.56 & +5.64 & -3.19 & 2.90 & $\pm$0.50 & 1.01 & +0.27 & -0.06 & \nodata \\
5695396 & 283 & 17.21 & 22.10 & +4.88 & -4.85 & 2.30 & $\pm$0.50 & 1.00 & +0.04 & -0.03 & 51.1 \\
2692377 & 299 & 23.43 & 24.86 & +6.61 & -6.58 & 1.90 & $\pm$0.50 & 0.93 & +0.04 & -0.03 & 90. \\
6063220 & 305 & 14.99 & 19.49 & +5.18 & -5.19 & 1.90 & $\pm$0.50 & 0.73 & +0.03 & -0.03 & 50.3 \\
6071903 & 306 & 17.84 & 24.25 & +6.80 & -6.78 & 1.80 & $\pm$0.50 & 0.86 & +0.03 & -0.03 & 47.4 \\
9139084 & 323 & 7.97 & 13.33 & +2.03 & -3.75 & 3.30 & $\pm$0.50 & 0.87 & +0.01 & -0.21 & 36.7 \\
10616571 & 340 & 12.93 & 8.84 & +1.91 & -0.93 & 6.60 & $\pm$0.50 & 1.15 & +0.23 & -0.08 & \nodata \\
10878263 & 341 & 19.21 & 21.77 & +5.50 & -6.95 & 2.20 & $\pm$0.50 & 0.95 & +0.10 & -0.21 & 90. \\
11074541 & 345 & 34.80 & 14.43 & +3.47 & -3.46 & 2.10 & $\pm$0.50 & 0.60 & +0.02 & -0.02 & \nodata \\
11100383 & 346 & 14.99 & 12.35 & +3.58 & -2.40 & 2.60 & $\pm$0.50 & 0.63 & +0.14 & -0.02 & 90. \\
6471021 & 372 & 11.90 & 13.57 & +4.67 & -2.43 & 2.80 & $\pm$0.50 & 0.75 & +0.22 & -0.01 & 90. \\
3323887 & 377 & 16.55 & 23.36 & +5.35 & -5.35 & 2.20 & $\pm$0.50 & 1.02 & +0.03 & -0.03 & 45.1 \\
4827723 & 632 & 17.43 & 25.06 & +7.32 & -8.74 & 1.80 & $\pm$0.50 & 0.89 & +0.08 & -0.19 & 44.1 \\
6707835 & 666 & 42.72 & 19.82 & +3.85 & -3.84 & 2.60 & $\pm$0.50 & 1.02 & +0.03 & -0.02 & \nodata \\
7115785 & 672 & 30.55 & 17.92 & +11.65 & -4.27 & 2.10 & $\pm$0.50 & 0.74 & +0.45 & -0.01 & \nodata \\
7630229 & 683 & 16.55 & 20.16 & +5.32 & -3.87 & 2.70 & $\pm$0.50 & 1.08 & +0.20 & -0.05 & 90. \\
9702072 & 714 & 28.25 & 27.04 & +8.26 & -10.28 & 1.70 & $\pm$0.50 & 0.91 & +0.08 & -0.22 & 90. \\
7825899 & 896 & 12.29 & 14.88 & +2.68 & -2.67 & 2.80 & $\pm$0.50 & 0.82 & +0.02 & -0.02 & 90. \\
1161345 & 984 & 8.60 & 10.07 & +1.12 & -1.13 & 4.60 & $\pm$0.50 & 0.92 & +0.02 & -0.02 & 90. \\
7295235 & 987 & 19.72 & 18.78 & +3.97 & -5.07 & 2.40 & $\pm$0.50 & 0.89 & +0.03 & -0.15 & 90. \\
2302548 & 988 & 12.29 & 14.86 & +2.78 & -4.35 & 2.70 & $\pm$0.50 & 0.79 & +0.02 & -0.18 & 55.8 \\
6362874 & 1128 & 33.26 & 19.53 & +4.34 & -5.24 & 2.30 & $\pm$0.50 & 0.89 & +0.04 & -0.14 & \nodata \\
10350571 & 1175 & 41.54 & 23.07 & +5.56 & -3.34 & 3.80 & $\pm$0.50 & 1.73 & +0.35 & -0.10 & \nodata \\
11336883 & 1445 & 5.13 & 3.98 & +0.56 & -0.25 & 15.50 & $\pm$0.50 & 1.22 & +0.17 & -0.06 & \nodata \\
\enddata
\tablenotetext{a}{\citet{2012ApJ...756...66H} points out close companions and likely contamination of this source}
\tablenotetext{b}{Also measured by \citet{2011ApJS..197...14D}}
\tablenotetext{c}{KOI 244 is has a high T$_{eff}$ \citep[6100K as measured in][]{2012Natur.486..375B}  and so variability may not be caused by starspots.}
\tablenotetext{d}{Also measured by \citet{2012ApJ...756...66H}}

\end{deluxetable*}

\begin{deluxetable*}{cccccccccc}
\tablecaption{Systems with Possible Spin-Orbit Misalignment}
\tablenum{5}
\tabletypesize{\footnotesize}
\tablehead{\colhead{KID} & \colhead{KOI} & \colhead{P$_{phot}$} & \colhead{P$_{spec}$} & \colhead{$plus$ err} & \colhead{$min$ err} & \colhead{{\em i$_\star$}} & \colhead{$plus$ err} & \colhead{$min$ err}  & \colhead{Notes} \\ 
\colhead{} & \colhead{} & \colhead{(days)} & \colhead{(days)} & \colhead{(days)} & \colhead{(days)} & \colhead{(degrees)} & \colhead{(degrees)} & \colhead{(degrees)} & \colhead{}} 

\startdata
  11554435 & 63 & 5.39 & 12.44 & 1.66 & 1.68 & 25.7 & 6.9 & 3.2 & 6.31R$_e$\\
  2444412 & 103 & 10.71 & 20.87 & 4.76 & 6.66 & 30.9 &16.3 & 6.2 & 2.95R$_e$\\
  5383248 & 261\tablenotemark{a} & 15.30 & 20.98 & 4.61 & 6.05 & 46.8 & 8.1 & 10.1 &  2.65R$_e$\\
  5695396 & 283 & 17.21 & 22.10 & 4.88 & 4.85 & 51.1 & 20.9 & 11.5 & FP + 0.85R$_e$\\
  6063220 & 305 & 14.99 & 19.49 & 5.18 & 5.19 & 50.3 & 12.5 & 12.9 & 1.57R$_e$ \\
  6071903 & 306 & 17.84 & 24.25 & 6.80 & 6.78 & 47.4 & 17.7 & 12.3 & 2.29R$_e$\\
  9139084 & 323 & 7.97 & 13.33 & 2.03 & 3.75 & 36.7 & 15.2 & 5.5 & 2.20R$_e$ \\
  3323887 & 377\tablenotemark{b} & 16.55 & 23.36 & 5.35 & 5.35 & 45.1 & 10.5 & 9.9 & 8.28R$_e$,8.22R$_e$,1.67R$_e$ \\
  4827723 & 632 & 17.43 & 25.06 & 7.32 & 8.74 & 44.1 & 20.4 & 11.5 & 1.46R$_e$\\
  2302548 & 988 & 12.29 & 14.86 & 2.78 & 4.35 & 55.8 & 17.8 & 11.6 & 2.21R$_e$; 2.17R$_e$ \\
 \enddata
\tablenotetext{a}{Also measured by \citet{2012ApJ...756...66H}}
\tablenotetext{b}{Kepler-9; \citet{2010Sci...330...51H}} 
\label{spinorbittable}
\end{deluxetable*}

\end{document}